\documentclass[11pt]{article}

\usepackage[T1]{fontenc}
\usepackage[utf8]{inputenc}

\usepackage[tt=false,type1=true]{libertine}
\usepackage[varqu]{zi4}
\usepackage[libertine]{newtxmath}
\usepackage{microtype}

\usepackage[
  a4paper,
  margin=1in,
  headheight=14pt,
  headsep=24pt,
  footskip=30pt
]{geometry}

\usepackage{fancyhdr}
\usepackage{amsmath}
\usepackage{booktabs}
\usepackage{graphicx}
\usepackage{url}
\usepackage{hyperref}
\usepackage{xcolor}
\usepackage{lscape}
\usepackage{float}
\usepackage{makecell}
\usepackage{seqsplit}
\usepackage{array}

\usepackage{booktabs}
\usepackage{tabularx}
\usepackage{longtable}
\usepackage{array}
\usepackage{multirow}
\usepackage{enumitem}
\usepackage{xcolor}
\usepackage{graphicx}
\usepackage{tikz}
\usetikzlibrary{arrows.meta,positioning,fit,shapes.geometric,shapes.multipart,calc}
\usepackage{pgfplots}
\pgfplotsset{compat=1.18}
\usepackage{hyperref}
\usepackage[nameinlink,noabbrev]{cleveref}
\usepackage{listings}
\usepackage{siunitx}
\usepackage{caption}
\usepackage{subcaption}
\usepackage{xspace}

\newcolumntype{Y}{>{\raggedright\arraybackslash}X}
\newcolumntype{L}[1]{>{\raggedright\arraybackslash}p{#1}}

\usepackage{makecell}
\usepackage{siunitx}

\usepackage[backend=biber,style=numeric]{biblatex}
\makeatletter

\newcommand{\@toptitlebar}{%
  \hrule height 2pt
  \vskip 0.25in
  \vskip -\parskip
}

\newcommand{\@bottomtitlebar}{%
  \vskip 0.29in
  \vskip -\parskip
  \hrule height 2pt
  \vskip 0.09in
}

\renewcommand{\maketitle}{%
  \par
  \begingroup
    \thispagestyle{plain}
    \begin{center}
      \vspace*{0.1in}
      \@toptitlebar
      {\LARGE\scshape \@title\par}
      \@bottomtitlebar
      \vspace{0.15in}
      {\normalsize\bfseries \@author\par}
      \vspace{0.2in}
    \end{center}
  \endgroup
}

\renewcommand{\section}{%
  \@startsection{section}{1}{\z@}%
                {-2.0ex \@plus -0.5ex \@minus -0.2ex}%
                {1.2ex \@plus 0.2ex}%
                {\large\bfseries\raggedright}}

\renewcommand{\subsection}{%
  \@startsection{subsection}{2}{\z@}%
                {-1.8ex \@plus -0.5ex \@minus -0.2ex}%
                {0.8ex \@plus 0.2ex}%
                {\normalsize\bfseries\raggedright}}

\renewcommand{\subsubsection}{%
  \@startsection{subsubsection}{3}{\z@}%
                {-1.5ex \@plus -0.5ex \@minus -0.2ex}%
                {0.5ex \@plus 0.2ex}%
                {\normalsize\bfseries\raggedright}}

\makeatother

\newcommand{\bench}{PQ-TLS Observability Benchmark v1\xspace}
\newcommand{\sigmap}{$\Sigma_P$\xspace}
\newcommand{\sigmaa}{$\Sigma_A$\xspace}
\newcommand{\sigmac}{$\Sigma_C$\xspace}
\newcommand{\sigmar}{$\Sigma_R$\xspace}
\newcommand{\Bzero}{B0\xspace}
\newcommand{\Bone}{B1\xspace}
\newcommand{\Btwo}{B2\xspace}
\newcommand{\Bthree}{B3\xspace}
\newcommand{\unknown}{\texttt{unknown}\xspace}
\newcommand{\ambiguous}{\texttt{ambiguous}\xspace}
\newcommand{\na}{\texttt{not\_applicable}\xspace}
\newcommand{\hybrid}{\texttt{X25519MLKEM768}\xspace}

\title{Observability for Post-Quantum TLS Readiness:\\
	A Multi-Surface Evidence Framework}

\author{%
José Luis Delgado \\
\small Universitat Oberta de Catalunya \\
\texttt{jdelgado13@uoc.edu}
}

\begin{document}

\maketitle

\begin{abstract}
	Post-quantum migration in Transport Layer Security (TLS) requires measurements that separate endpoint capability, session negotiation, certificate-chain evidence, and the provenance of missing observations. Reliable assessment must account for protocol opacity, measurement limits, and temporal drift, especially under TLS 1.3 encryption, resumption, mutual TLS, trace truncation, fragmentation, coalescing, and active certificate retrieval.
	
	We present a multi-surface framework for post-quantum TLS observability. The framework separates passive session evidence, active probing, certificate-chain evidence, and registry knowledge, and maps these sources onto measurement planes for session behavior, key establishment, endpoint capability, authentication, lifecycle, observability, and policy. We instantiate the framework as a reproducible artifact with schema-enforced observations and results, versioned registries, auditable inference rules, stress contracts and baseline adapters.
	
	We evaluate the framework on 29 controlled scenarios covering TLS 1.2 and TLS 1.3, classical and hybrid key establishment, mutual TLS, resumption, HelloRetryRequest, truncation, fragmentation and coalescing, temporal drift, IPv6, and chain-depth variation. Passive evidence closes session-level planes, active probing establishes capability lower bounds, and multi-surface evidence closes the full measurement object. Stress scenarios exercise cases where sound inference preserves uncertainty or surfaces contradiction. Against an inherited TLS quantum-vulnerability analyzer, the baseline detects 2 of 29 runs and 0 of 23 TLS 1.3 runs. In a stratified public campaign over 1000 targets and 2000 fresh probes, the framework completes 1971 handshakes, collects 1368 chain artifacts, confirms hybrid capability for 310 targets, and identifies 310 cases where endpoint capability exceeds what any single classical session view reveals. A repeated round preserves the same hybrid-confirmed count, shows high clear-complete stability, and exposes small capability and certificate drift.
	
	These results support a measurement model for post-quantum TLS readiness based on explicit evidence surfaces, per-plane closure, active corroboration, source linkage, and first-class treatment of \unknown, \na, ambiguity, and contradiction.
\end{abstract}


\section{Introduction}
\label{sec:introduction}

The transition from classical public-key cryptography to post-quantum cryptography (PQC) is now an operational concern for deployed Transport Layer Security (TLS). NIST has standardized ML-KEM for key encapsulation and ML-DSA and SLH-DSA for digital signatures~\cite{NISTFIPS203,NISTFIPS204,NISTFIPS205}. The TLS ecosystem is experimenting with, deploying, and standardizing hybrid key-establishment mechanisms that combine elliptic-curve Diffie-Hellman with ML-KEM, including variants such as X25519MLKEM768 for TLS 1.3~\cite{IETFECDHEMLKEM}. Operators, auditors, and defenders therefore need measurements that identify endpoint capability, session negotiation, certificate-chain cryptographic material, and the evidentiary limits of incomplete observations.

We frame post-quantum TLS assessment as an observability problem. TLS exposes cryptographic posture through partial, surface-dependent evidence. TLS 1.3 encrypts more handshake material than TLS 1.2~\cite{RFC8446,RFC5246}. Certificates visible in passive TLS 1.2 captures are usually unavailable in ordinary passive TLS 1.3 captures. Active probing can reveal endpoint behavior, but it produces evidence on a different plane from a concrete captured session. A server may support a hybrid group while a particular client negotiates a classical group. A certificate chain collected after a captured flow may describe the endpoint at retrieval time while having limited linkage to the original session. PSK resumption, HelloRetryRequest, mTLS, temporal drift, middleboxes, truncation, and IPv6 handling add further seams between observation and inference. Faithful measurement requires source-aware evidence, explicit linkage, and per-field closure.

Recent work has begun to automate packet-level analysis of quantum-vulnerable TLS. Cho et al. present an open-source framework for automated analysis of quantum-vulnerable TLS through packet inspection, including hierarchical packet filtering, TLS parsing, detection of classical and hybrid configurations, and a certificate acquisition strategy for encrypted TLS 1.3 certificates~\cite{Cho2026CryptoAgilityTLS}. Their work moves PQC assessment toward operational visibility. For TLS 1.3 certificates, the implementation initiates a new TLS connection using the extracted SNI and retrieves the certificate through OpenSSL. This produces useful endpoint evidence and illustrates a recurring measurement pattern: one surface supplies a session observation, another surface supplies corroborating endpoint or certificate evidence. The resulting measurement object must record what packet inspection supports, what active probing adds, how endpoint support differs from negotiated parameters, how certificate evidence links to session evidence, and how errors, \unknown, \na, ambiguity, and contradiction propagate into conclusions.

We present a measurement framework and artifact for post-quantum TLS observability. The framework separates evidence into four surfaces: passive session evidence (\sigmap), active probe evidence (\sigmaa), certificate-chain evidence (\sigmac), and registry/rule evidence (\sigmar). It projects these observations onto seven planes: session, key establishment, capability, authentication, lifecycle, observability, and policy. The central output is a measurement object with explicit provenance, source linkage, ambiguity, and policy projection.

We build the benchmark and evaluation methodology around this measurement model. The controlled benchmark, \bench, contains canonical scenarios with strong ground truth and stress scenarios evaluated by behavioral contracts. Canonical scenarios test extraction under clean evidence. Stress scenarios exercise incomplete, contradictory, or non-observable evidence. In a truncated TLS 1.3 capture, \unknown can be the sound result for selected fields; a complete-looking label without supporting evidence is an inference error.

The accompanying reproducibility artifact is publicly available at \url{https://github.com/hypergalois/pqc-tls-observability}.

We evaluate the artifact in two settings. First, we compare an inherited baseline, based on the reference TLS quantum-vulnerability analyzer, against passive-only, passive-active, and full multi-surface modes on the same 29-scenario local benchmark slice. Second, we run a stratified public campaign over 1000 targets and a 250-target baseline-zoo slice comparing SSLyze, testssl.sh, single-probe modes, and dual-probe-plus-chain measurement. The evaluation connects scanner-level outcomes to the evidence available to each mode. It shows how hybrid support, classical negotiation, certificate evidence, temporal drift, and incomplete observations require a multi-surface measurement object with explicit uncertainty handling.

\paragraph{Contributions.}

\begin{enumerate}[leftmargin=1.4em]
	\item \textbf{A multi-surface evidence model for post-quantum TLS observability.} We define separate evidence surfaces for passive session evidence, active probe evidence, certificate-chain evidence, and registry knowledge. The model separates evidence from policy judgments and maps observations to session, key-establishment, capability, authentication, lifecycle, observability, and policy planes.
	
	\item \textbf{A schema-enforced reproducible artifact.} We instantiate the model as an auditable measurement stack with JSON Schemas for scenarios, observations, ground truth, target inventories, and inferred results; versioned registries for TLS groups, signature schemes, X.509 SPKI OIDs, X.509 signature OIDs, and aliases; rule files for inference, policy, and stress contracts; and adapters for inherited and external baselines.
	
	\item \textbf{A reproducible benchmark for PQ-TLS observability.} We construct \bench with 29 executed scenarios: 14 canonical and 15 stress scenarios covering TLS 1.2, TLS 1.3, classical and hybrid key establishment, mTLS, resumption, HRR, truncation, fragmentation/coalescing, temporal drift, IPv6, and chain-depth variation.
	
	\item \textbf{A contract-based stress evaluation regime.} Stress scenarios require contract-based evaluation alongside exact-match ground truth. Some scenarios are designed to produce ambiguity, contradiction, or \na states. We encode these expectations as explicit behavioral contracts and reward correct uncertainty handling alongside field extraction.
	
	\item \textbf{A baseline comparison across measurement modes.} We compare an inherited packet-inspection baseline (\Bzero) with passive-only (\Bone), passive-active (\Btwo), and multi-surface (\Bthree) modes on the same benchmark slice. \Bzero detects 2/29 runs and 0/23 TLS 1.3 runs, while the multi-surface stack retains broad closure across planes.
	
	\item \textbf{A stratified public measurement study.} We run a 1000-target, 2000-probe public campaign and a 250-target scanner comparison slice. The dual-probe measurement mode confirms hybrid capability for targets outside the hybrid set reported by classical scanner baselines under the configured budget. A repeated public round confirms the stability of the main hybrid-capability result.
	
	\item \textbf{Operational guidance for crypto-agility auditing.} We distill when passive evidence closes a measurement plane, when active corroboration is needed, when chain evidence closes authentication and lifecycle, and when uncertainty or contradiction should remain first-class measurement outputs.
\end{enumerate}

\paragraph{Research questions.}
The work is organized around five research questions, shown in \Cref{tab:research-questions}. The goal is to determine which evidence supports which statement about the endpoint or session.

\begin{table}[t]
\centering
\caption{Research questions and corresponding evidence.}
\label{tab:research-questions}
\begin{tabularx}{\textwidth}{@{}l X X@{}}
\toprule
RQ & Question & Evidence used \\
\midrule
RQ1 & Which post-quantum TLS properties can be inferred from passive session evidence alone? & Canonical and stress results for B1, TLS 1.2 visible certificates, and TLS 1.3 hidden authentication cases. \\
RQ2 & What does active probing add beyond passive observation? & B2 capability closure, support-vs-negotiation stress cases, and the public dual-probe campaign. \\
RQ3 & When does certificate-chain evidence close authentication and lifecycle, and when does linkage weaken the claim? & B3 results, chain-depth sentinel, temporal-drift stress contracts, and public chain artifacts. \\
RQ4 & How should incomplete, contradictory, or non-applicable evidence be scored? & Stress-contract evaluation, object-complete-clear metric, and explicit unknown/not-applicable states. \\
RQ5 & Do common scanner baselines reveal hybrid support under the same public slice and budget? & 250-target baseline-zoo comparison with SSLyze, testssl.sh, single-probe, dual-probe, and chain modes. \\
\bottomrule
\end{tabularx}
\end{table}

The remainder of the manuscript proceeds as follows. \Cref{sec:background} reviews TLS,
post-quantum migration, crypto-agility inventory, and the limits of packet-inspection approaches.
\Cref{sec:model} defines the multi-surface measurement model, evidence planes, measurement
object, and unresolved-state semantics. \Cref{sec:system} describes the system architecture,
artifact design, registries, inference procedure, measurement modes, and stress contracts.
\Cref{sec:benchmark} presents the controlled benchmark design, including canonical and stress
scenario families. \Cref{sec:methodology} details the controlled execution, baseline execution,
public campaign profiles, repeated rounds, and statistical treatment. \Cref{sec:benchmark-results}
reports the controlled benchmark results and inherited-baseline comparison. \Cref{sec:public-study}
analyzes the stratified public measurement campaign, repeated-round stability, family
heterogeneity, and probing-mode effects. \Cref{sec:baseline-zoo} compares scanner and probe
baselines on the stratified slice. \Cref{sec:discussion} discusses the implications for separating
support, negotiation, certificate evidence, uncertainty, and operational readiness. We then review
related work, state limitations and threats to validity, describe reproducibility and artifact
details, and conclude.

\section{Background}
\label{sec:background}

\subsection{TLS 1.2, TLS 1.3, and cryptographic negotiation}

TLS provides confidentiality, integrity, and authentication for network communication. TLS 1.2 and TLS 1.3 expose different measurement surfaces. In TLS 1.2, the cipher-suite name encodes a bundle of cryptographic choices, including key exchange, authentication, symmetric encryption, and integrity protection~\cite{RFC5246}. With a complete passive capture, the server certificate chain is visible in the handshake.

TLS 1.3 changes both the protocol design and the available passive evidence. It removes many legacy mechanisms, reduces round trips, moves key-exchange negotiation into extensions such as \texttt{supported\_groups} and \texttt{key\_share}, and encrypts additional handshake messages, including the certificate message sent after \texttt{ServerHello}~\cite{RFC8446}. A TLS 1.3 cipher suite such as \texttt{TLS\_AES\_128\_GCM\_SHA256} identifies the AEAD and hash/KDF choices; the selected key-establishment group must be measured separately. The server certificate chain is also usually unavailable in ordinary passive TLS 1.3 capture, so authentication evidence often requires another collection surface.

\subsection{Post-quantum transition and hybrid TLS}

The quantum threat to deployed public-key cryptography follows from Shor's algorithm, which breaks the mathematical assumptions underlying RSA, finite-field Diffie-Hellman, and elliptic-curve cryptography given a sufficiently capable quantum computer~\cite{Shor1994,RSA1978,Koblitz1987ECC}. Symmetric cryptography has a different risk profile: Grover's algorithm gives a quadratic search speedup, motivating larger symmetric security margins without the same structural break~\cite{Grover1996}.

NIST's PQC standardization process has produced ML-KEM for key encapsulation and ML-DSA and SLH-DSA for signatures~\cite{NISTFIPS203,NISTFIPS204,NISTFIPS205}. In TLS, the near-term confidentiality migration path has centered on hybrid key establishment. Hybrid designs combine a classical ECDHE shared secret with an ML-KEM shared secret, so that the resulting session keys depend on both components~\cite{IETFECDHEMLKEM,IETFHybridDesign}. Measuring hybrid TLS requires separate evidence for endpoint support and per-session negotiation. Client profile, server preference, HelloRetryRequest behavior, and middlebox constraints can all affect the group selected in a concrete flow.

Authentication is a separate migration plane. Replacing or augmenting RSA/ECDSA certificates with post-quantum signatures introduces chain-size, compatibility, and PKI deployment considerations. Drafts for ML-DSA use in TLS 1.3 and composite ML-DSA illustrate the direction of this transition~\cite{IETFMLDSA,IETFCompositeMLDSA}. Separate fields for key establishment and authentication preserve which part of the TLS stack has migrated and which evidence supports that assessment.

\subsection{Crypto-agility as an inventory problem}

Crypto-agility is commonly described as the ability to replace cryptographic algorithms without disrupting deployed systems. Operationally, migration begins with inventory. Operators need to identify algorithms, dependencies, protocol contexts, and deployment points before choosing migration actions. NIST guidance on cryptographic agility emphasizes the identification of algorithms, dependencies, and deployment contexts as a prerequisite for migration planning~\cite{NISTCryptoAgility2025}.

For TLS, inventory spans configuration, negotiation, certificates, and deployment topology. Public services may be fronted by load balancers, CDNs, regional endpoints, middleboxes, and certificate automation. A service may advertise different groups to different client profiles or select different parameters across connection attempts. Measurement methodology therefore becomes part of crypto-agility: the inventory must record what was observed, where it was observed, and which inference each observation supports.

\subsection{The gap in packet-inspection approaches}

Packet inspection is useful because it observes parameters negotiated in concrete sessions. Cho et al. provide a reference point for automated packet-level analysis of quantum-vulnerable TLS, with packet filtering, TLS handshake parsing, detection of classical and hybrid configurations, and certificate retrieval for TLS 1.3 through a separate SNI-based connection~\cite{Cho2026CryptoAgilityTLS}. This certificate retrieval strategy supplies endpoint evidence for encrypted TLS 1.3 handshakes and highlights the need to separate evidence surfaces.

A certificate chain fetched through a new active connection describes the endpoint at probe time. Its linkage to an earlier captured session depends on server configuration, certificate rotation, regional routing, client-profile variation, and timing. These conditions can make passive session evidence, active probe evidence, and certificate-chain evidence diverge. A faithful measurement object therefore records the source of each observation, the plane it supports, the linkage between surfaces, and the status of incomplete, ambiguous, contradictory, or not-applicable fields.

\section{Measurement Model}
\label{sec:model}

This section defines the measurement model used throughout the paper. The model keeps observation, inference, policy, evidence, and confidence as separate components.

\subsection{Evidence surfaces}

We define four evidence surfaces, summarized in \Cref{tab:surfaces}.

\begin{table}[t]
\centering
\caption{Evidence surfaces used by the measurement model.}
\label{tab:surfaces}
\begin{tabularx}{\textwidth}{@{}l l X@{}}
\toprule
Surface & Name & Examples of evidence \\
\midrule
$\Sigma_P$ & Passive session & TLS record and handshake fields, negotiated version, cipher suite, selected group when visible, HRR, completeness, TLS 1.2 certificate visibility, packet truncation and fragmentation flags. \\
$\Sigma_A$ & Active probe & Client probe profile, observed probe negotiation, multi-probe lower-bound capability inference, same-run hidden session detail such as TLS 1.3 mTLS. \\
$\Sigma_C$ & Certificate chain & Leaf/intermediate/root chain artifacts, SPKI and signature algorithms, validity interval, chain depth, chain profile, chain-source linkage. \\
$\Sigma_R$ & Registry and rules & Named-group registry, signature-scheme registry, X.509 OIDs, aliases, draft/final/obsolete status, inference rules, and policy rules. \\
\bottomrule
\end{tabularx}
\end{table}

\sigmap is passive session evidence. It includes fields extracted from packet capture: TLS version negotiation, selected group when visible, cipher suite, HelloRetryRequest visibility, session completion, TLS 1.2 certificate visibility, and capture completeness. It provides the strongest evidence for the behavior of a concrete observed flow. In TLS 1.3, its coverage is limited by encrypted handshake material.

\sigmaa is active evidence. It comes from probes executed by the measurement artifact. We split it into three subfields: \texttt{probe\_profile}, describing what the client offered; \texttt{observed\_negotiation}, describing what the probe negotiated; and \texttt{endpoint\_capability\_inference}, describing the lower bound inferred from one or more probes. These fields preserve the difference between client offer, observed negotiation, and inferred server capability.

\sigmac is certificate-chain evidence. It includes chain artifacts, leaf and intermediate algorithms, signature algorithms, validity windows, chain depth, and lifecycle categories. A chain may be visible in TLS 1.2 passive capture, collected by an active probe, or loaded from a controlled scenario artifact. Each source receives an explicit linkage label.

\sigmar is registry evidence. It maps raw identifiers to canonical names, tracks aliases, distinguishes final, draft, obsolete, experimental, and vendor-specific identifiers, and stores rule provenance. Post-quantum TLS identifiers are still changing, so the registry preserves historical draft aliases alongside final identifiers and retains deployment history.

\subsection{Measurement planes}

The evidence surfaces project onto measurement planes. \Cref{tab:planes} summarizes the planes and their closure conditions.

\begin{table}[t]
\centering
\caption{Measurement planes and closure conditions.}
\label{tab:planes}
\begin{tabularx}{\textwidth}{@{}l l X@{}}
\toprule
Plane & Typical source & Closure condition \\
\midrule
Session core & $\Sigma_P$ & Version, selected group or justified non-applicability, completeness, and core handshake state are known. \\
Session hidden detail & $\Sigma_A$ or linked evidence & Encrypted or otherwise hidden session details, such as TLS 1.3 client authentication, are closed only through explicit linkage. \\
Key establishment & $\Sigma_P,\Sigma_R$ & Classical, hybrid, PQ, or not-applicable key-establishment profile is resolved from selected group or protocol semantics. \\
Capability & $\Sigma_A$ & A lower bound on endpoint support is inferred from one or more probe negotiations under declared client profiles. \\
Authentication & $\Sigma_P$ or $\Sigma_C$ & Leaf and chain authentication profile are resolved from visible or collected certificate evidence. \\
Lifecycle & $\Sigma_C$ & Certificate validity interval and short/long-lived bucket are known. \\
Observability & all surfaces & Linkage, confidence, ambiguity, contradiction, and source provenance are recorded. \\
Policy & measurement object & Optional downstream projection maps the measurement object to operator-facing verdicts. \\
\bottomrule
\end{tabularx}
\end{table}

A measurement object can be closed in one plane and open in another. A passive TLS 1.3 capture may close session core and key establishment while leaving authentication and lifecycle unresolved. A dual-probe campaign can close a lower bound on endpoint capability. A chain probe can close authentication for the probed endpoint at a given time, with linkage recorded separately for any earlier encrypted session.

\subsection{Measurement object}

A measurement result is represented as a structured object:
\begin{equation}
	M = \langle S, K, C, A, L, O, P \rangle,
\end{equation}
where $S$ is the session profile, $K$ is the key-establishment profile, $C$ is the capability profile, $A$ is the authentication profile, $L$ is the lifecycle profile, $O$ is the observability/linkage profile, and $P$ is an optional policy projection.

The policy projection is evaluated last. For example, a session profile may state that a session negotiated X25519, while a capability profile states that a hybrid probe negotiated X25519MLKEM768. A policy profile may classify that endpoint as ``hybrid-capable with classical negotiation under the default client''. That statement assigns the classical result to the captured session and the hybrid result to endpoint capability. This separation gives operators a precise account of which evidence supports each claim.

\subsection{Unknown, ambiguous, contradictory, and not applicable}

The artifact represents unresolved and exceptional states as first-class outcomes. We distinguish:

\begin{itemize}[leftmargin=1.4em]
	\item \unknown: evidence is insufficient to decide a property
	\item \ambiguous: evidence exists but is not uniquely interpretable
	\item contradictory: two or more evidence surfaces disagree in a way that should remain visible
	\item \na: the property does not apply to the scenario, for example a TLS 1.2 static-RSA session has no selected ECDHE or hybrid named group
\end{itemize}

The \unknown and \na states serve different measurement roles. A TLS 1.2 static-RSA session receives \na for selected ECDHE or hybrid named group because the property is outside the scenario semantics. Hidden TLS 1.3 client-auth detail receives \unknown when the capture lacks sufficient evidence. The artifact separates session core from hidden session detail and records linkage per plane.

\section{System and Artifact Design}
\label{sec:system}

\subsection{Design requirements}

The architecture follows six design requirements, summarized in \Cref{tab:design-requirements}. These requirements serve as methodological safeguards for evidence-structured measurement. R1 and R2 keep active endpoint evidence distinct from passive session evidence. R3 keeps policy projection separate from raw measurement. R4 and R5 make unresolved and contradictory evidence explicit evaluation outcomes with their own scoring semantics.

\begin{table}[t]
\centering
\caption{Design requirements derived from the measurement model.}
\label{tab:design-requirements}
\begin{tabularx}{\textwidth}{@{}l X X@{}}
\toprule
Requirement & Rationale & Artifact mechanism \\
\midrule
R1: Separate surfaces & Passive, active, and chain evidence answer different questions. & Independent observation streams \sigmap, \sigmaa, \sigmac and registry \sigmar. \\
R2: Preserve provenance & Active enrichment can be useful while remaining distinct from passive session evidence. & Timestamps, source artifacts, chain-source types, and per-plane linkage. \\
R3: Separate measurement and policy & Readiness profiles may change while raw measurements remain valid. & Measurement object first, policy projection second. \\
R4: Support non-positive states & Unknown and not-applicable outcomes are sometimes correct. & Explicit \unknown, \ambiguous, contradiction, and \na states. \\
R5: Use stress contracts & Exact-match scoring misrepresents intentionally unresolved scenarios. & Contract-based stress evaluation. \\
R6: Version identifiers & PQ-TLS identifiers evolve across drafts, vendors, and standards. & Versioned registry with status and alias metadata. \\
\bottomrule
\end{tabularx}
\end{table}

\subsection{Architecture}

\begin{figure}[t]
	\centering
	\begin{tikzpicture}[
    node distance=1.1cm and 1.0cm,
    box/.style={draw, rounded corners, align=center, minimum width=2.8cm, minimum height=0.9cm, font=\small},
    wide/.style={draw, rounded corners, align=center, minimum width=7.0cm, minimum height=0.9cm, font=\small},
    arrow/.style={-Latex, thick}
]

\node[box, fill=gray!8] (passive) {$\Sigma_P$\\Passive session\\pcap/handshake};
\node[box, fill=gray!8, right=of passive] (active) {$\Sigma_A$\\Active probe\\profiles};
\node[box, fill=gray!10, right=of active] (chain) {$\Sigma_C$\\Certificate chain\\artifacts};
\node[box, fill=gray!8, right=of chain] (registry) {$\Sigma_R$\\Registry/rules\\aliases};

\node[wide, below=1.2cm of active, xshift=1.7cm, fill=gray!8] (engine) {Rule engine: canonicalization, linkage, ambiguity, contradiction};
\node[wide, below=1.1cm of engine, fill=gray!10] (measurement) {Measurement object: session, key establishment, capability, authentication, lifecycle, observability};
\node[wide, below=1.1cm of measurement, fill=gray!6] (policy) {Optional policy projection: readiness verdicts, guidance, dashboards};

\draw[arrow] (passive) -- (engine);
\draw[arrow] (active) -- (engine);
\draw[arrow] (chain) -- (engine);
\draw[arrow] (registry) -- (engine);
\draw[arrow] (engine) -- (measurement);
\draw[arrow] (measurement) -- (policy);

\end{tikzpicture}
	\caption{Multi-surface architecture. Passive session evidence, active probe evidence, certificate-chain evidence, and registry knowledge remain separate until the rule engine constructs a measurement object. Policy projection is downstream of measurement.}
	\label{fig:architecture}
\end{figure}
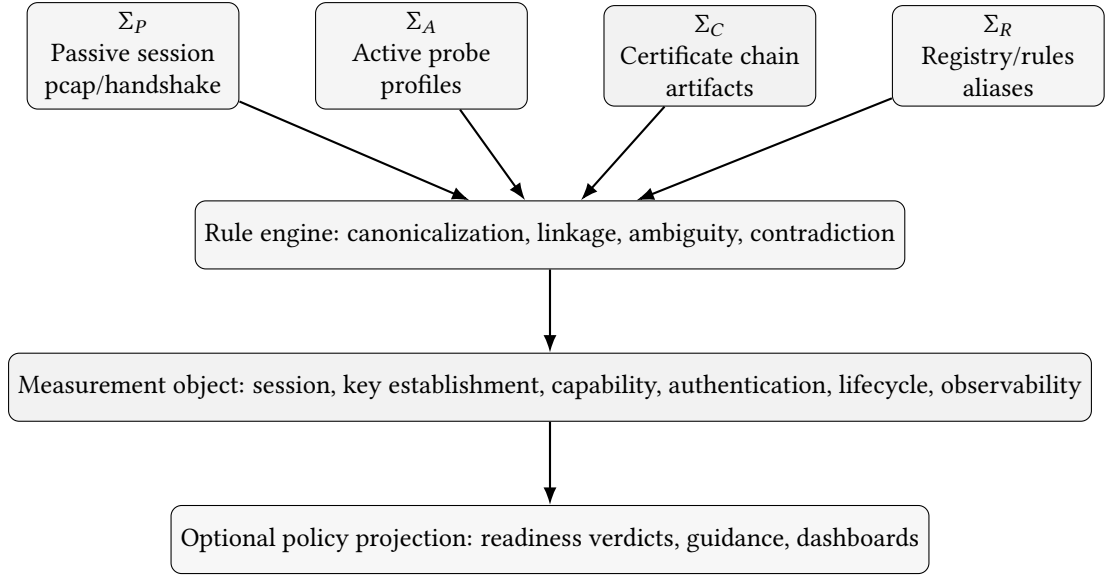

\Cref{fig:architecture} shows the system architecture. The artifact is organized as a measurement pipeline. Raw evidence is extracted from passive captures, active probes, and certificate-chain collection. The registry normalizes identifiers. The rule engine emits measurement objects. Policy profiles consume completed measurement objects as downstream projections.

The repository contains explicit JSON Schemas for observations, scenarios, ground truth, target inventories, and results. JSON/JSONL is the canonical machine-readable format, CSV is used for imported legacy input and derived interchange. Each observation records provenance, timestamps, parser version, registry version, source artifact, and ambiguity flags.

\subsection{Artifact implementation and verifiability}

The framework is implemented as an auditable measurement stack. The repository separates schemas, registries, parsers, inference rules, policy rules, baseline adapters, evaluation scripts, benchmark corpus metadata, campaign runners, and derived results.

The artifact defines JSON Schemas for scenarios, observations, ground truth, target inventories, and inferred results. Each observation records an evidence surface, parser and registry versions, a source artifact, SHA-256 digest, capture timestamps, raw fields, canonical fields, completeness flags, ambiguity status, and provenance. Each inferred result is schema-constrained to include session, capability, key-establishment, authentication, lifecycle, observability, policy, inference trace, ambiguity, and contradiction fields.

The registry bundle contains named groups, signature schemes, X.509 SPKI OIDs, X.509 signature OIDs, and aliases. Registry entries preserve identifier status and keep historically distinct post-quantum identifiers separate, including obsolete Kyber draft names and final ML-KEM hybrid names. The rule layer is external to parser code and includes inference rules, policy profiles, and stress contracts. This separation supports independent audit of extraction code, measurement semantics, and policy projection.

The artifact also includes adapters for the inherited \Bzero analyzer, SSLyze, and testssl.sh; a controlled benchmark corpus; public-campaign inventory schemas; active campaign runners; drift analysis; multiprobe analysis; verifier scripts; and paper-ready exports. The frozen benchmark release is identified as \texttt{pq\_tls\_observability}, version \texttt{1.0.0}. The artifact freeze includes verifier scripts for the benchmark release, \Bzero comparison, CCS extension, and baseline-zoo slice.

\subsection{Identifier registry}

The registry stores TLS named groups, signature schemes, X.509 SPKI OIDs, X.509 signature OIDs, and aliases. Each entry records a canonical name, raw identifiers, aliases, family, status, and source notes. The status field distinguishes \texttt{standard}, \texttt{draft}, \texttt{obsolete}, \texttt{experimental}, and \texttt{vendor}. The rule engine preserves distinctions among historically different identifiers, such as early Kyber draft groups and final ML-KEM hybrid groups. This keeps deployment history visible and avoids treating all post-quantum-looking identifiers as the same migration state.

\subsection{Modes B0 through B3}

We compare four modes, summarized in \Cref{tab:modes}.

\begin{table}[t]
\centering
\caption{Compared measurement modes.}
\label{tab:modes}
\begin{tabularx}{\textwidth}{@{}l l X@{}}
\toprule
Mode & Short name & Semantics \\
\midrule
B0 & Inherited baseline & Adapter around the reference TLS quantum-vulnerability analyzer. Outputs are normalized for comparison, while semantics remain detector-oriented. \\
B1 & Passive-only & Uses passive packet evidence and extracts TLS 1.2 certificate-derived auth/lifecycle when visible. It does not use active TLS 1.3 chain retrieval. \\
B2 & Passive-active & Adds active probing, multi-probe capability lower bounds, and same-run hidden-detail closure when explicitly linked. \\
B3 & Multi-surface & Combines passive, active, chain, and registry evidence to emit measurement objects with linkage and policy projection. \\
\bottomrule
\end{tabularx}
\end{table}

\Bzero is the inherited baseline around the reference TLS quantum-vulnerability analyzer. It provides the packet-level detector baseline for comparison with evidence-structured measurement.

\Bone is passive-only parsing. It reads packet evidence and, where visible, extracts TLS 1.2 certificate-derived authentication and lifecycle information. TLS 1.3 certificate retrieval through active connections is outside this mode.

\Btwo adds active probing. It can close endpoint capability lower bounds and same-run hidden details, including TLS 1.3 mTLS when explicitly linked.

\Bthree is the full multi-surface mode. It combines passive, active, chain, and registry evidence under the rule engine, with explicit linkage, ambiguity, contradiction, and policy separation.

\subsection{Inference procedure}

The inference procedure follows a fixed order. It first canonicalizes raw identifiers against the registry. It then builds per-surface observations. The rule engine infers plane profiles from those observations. Optional policy profiles consume the completed measurement object. A simplified pseudocode view is shown below.

\begin{lstlisting}[basicstyle=\ttfamily\small,breaklines=true]
	for scenario_or_target in inputs:
	obs_p = parse_passive_capture(scenario_or_target.pcap)
	obs_a = aggregate_active_probes(scenario_or_target.probes)
	obs_c = collect_or_load_chains(scenario_or_target.chain_artifacts)
	canon = canonicalize(obs_p, obs_a, obs_c, registry)
	
	result.session_profile = infer_session(canon.SigmaP, canon.SigmaA)
	result.key_establishment_profile = infer_key_establishment(result.session_profile, registry)
	result.capability_profile = infer_capability(canon.SigmaA)
	result.authentication_profile = infer_authentication(canon.SigmaP, canon.SigmaC)
	result.lifecycle_profile = infer_lifecycle(canon.SigmaC)
	result.observability_profile = infer_linkage_and_contradictions(canon)
	
	result.policy_projection = apply_policy_profiles(result)  # optional
\end{lstlisting}

Policy projection is evaluated after measurement. A hybrid key-establishment result supplies evidence for the key-establishment plane; certificate evidence must still be supplied by the passive or chain surface. A certificate chain obtained through an active probe can close the authentication plane for the endpoint at probe time. Its source remains active endpoint evidence, with linkage recorded separately for any captured session.

\subsection{Stress contracts}

Canonical scenarios are evaluated against exact-match ground truth. Stress scenarios are evaluated through contracts that specify the expected treatment of partial, conflicting, or asymmetric evidence. Truncated captures leave selected session fields unresolved. Temporal drift surfaces contradiction. Support-versus-negotiation cases mark endpoint capability as broader than the observed session. This evaluation regime scores outputs against evidence availability, so correctness includes explicit uncertainty, contradiction, and \na states when the scenario calls for them.

\section{Benchmark Design}
\label{sec:benchmark}

\subsection{Benchmark v1 overview}

The controlled laboratory artifact is \bench. It contains 29 executed scenarios: 14 canonical scenarios and 15 stress scenarios. \Cref{tab:benchmark-overview} summarizes the suite.

\begin{table}[t]
	\centering
	\caption{Benchmark v1 overview. Counts are executed laboratory scenarios.}
	\label{tab:benchmark-overview}
	\begin{tabularx}{\linewidth}{@{}l r Y@{}}
		\toprule
		Suite & Count & Evaluation regime and phenomena \\
		\midrule
		Canonical & 14 & Exact-match truth: TLS 1.2/1.3, classical/hybrid, RSA/ECDSA leaves, mTLS sentinels, lifecycle variants, IPv6, leaf+intermediate chain. \\
		Stress & 15 & Behavioral contracts: support vs negotiation, truncation, fragmentation/coalescing, temporal drift, HRR degradation, PSK/resumption, mTLS/resumption crossover. \\
		\midrule
		Total & 29 & 6 TLS 1.2 scenarios and 23 TLS 1.3 scenarios. \\
		\bottomrule
	\end{tabularx}
\end{table}

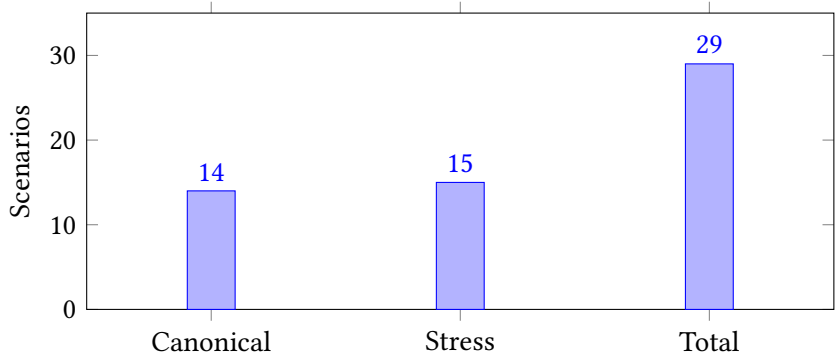
\begin{figure}[t]
	\centering
	\begin{tikzpicture}
\begin{axis}[
    ybar,
    width=0.72\textwidth,
    height=5.5cm,
    ymin=0, ymax=35,
    ylabel={Scenarios},
    symbolic x coords={Canonical,Stress,Total},
    xtick=data,
    nodes near coords,
    bar width=18pt,
    enlarge x limits=0.25,
]
\addplot coordinates {(Canonical,14) (Stress,15) (Total,29)};
\end{axis}
\end{tikzpicture}
	\caption{Benchmark v1 composition. The controlled benchmark separates canonical exact-match scenarios from stress-contract scenarios.}
	\label{fig:benchmark-overview}
\end{figure}

The canonical suite covers clean behavior across TLS versions, key-establishment families, authentication types, mTLS sentinels, lifecycle variants, IPv6, and chain depth. The stress suite covers support-versus-negotiation gaps, truncation, fragmentation/coalescing, temporal drift, HRR under degraded passive evidence, PSK resumption, and mTLS/resumption crossover.

\subsection{Canonical scenarios}

Canonical scenarios have strong ground truth derived from controlled configuration and artifacts. They include TLS 1.2 static RSA, TLS 1.2 ECDHE\_RSA, TLS 1.2 ECDHE\_ECDSA, a TLS 1.2 mTLS sentinel, TLS 1.3 classical X25519, TLS 1.3 hybrid X25519MLKEM768, RSA and ECDSA leaf certificates, TLS 1.3 mTLS sentinels, short-lived and long-lived certificate lifecycles, a leaf-plus-intermediate chain sentinel, and an IPv6 loopback sentinel.

The canonical suite evaluates whether the measurement object closes the expected planes when the relevant evidence is clean and available.

\subsection{Stress scenarios}

Stress scenarios exercise measurement behavior under partial, conflicting, asymmetric, or layout-sensitive evidence. The suite includes:

\begin{itemize}[leftmargin=1.4em]
	\item \textbf{Support versus negotiation:} classical and hybrid-capable probe profiles are applied to the same endpoint, exposing endpoint capability beyond the group selected in a single session.
	\item \textbf{Truncation:} captures are cut at different points, including before \texttt{ServerHello}, leaving selected session fields unresolved.
	\item \textbf{Fragmentation/coalescing:} packet layout is varied to exercise parsing across record and packet boundary changes.
	\item \textbf{Temporal drift:} passive session evidence is separated from later active or chain evidence, exposing configuration changes as contradictions across surfaces.
	\item \textbf{HRR degradation:} HelloRetryRequest is visible while the second leg is incomplete, leaving final negotiation unresolved.
	\item \textbf{PSK/resumption:} resumption scenarios exercise continuity semantics and the relationship between fresh authentication evidence and inherited session context.
	\item \textbf{mTLS crossover:} TLS 1.3 client-auth detail is hidden in passive capture and closed through same-run linked active evidence.
\end{itemize}

\subsection{Ground truth and provenance}

For laboratory scenarios, ground truth is generated from scenario configuration, server and client logs, certificate artifacts, and independent certificate oracles. The artifact records both OpenSSL-derived and asn1crypto-derived chain views and normalizes distinguished names before checking oracle consistency.

For public measurements, the evaluation uses confirmed lower bounds. A target is hybrid-confirmed when at least one controlled probe negotiates a hybrid group under a declared profile. The resulting claim is scoped to the configured probes, profiles, targets, and campaign vantage points.

\section{Experimental Methodology}
\label{sec:methodology}

\subsection{Controlled benchmark execution}

The controlled benchmark is executed through a local harness that launches configured TLS services and clients, captures passive traffic, records active probe logs, collects certificate chains, and emits observation objects. The harness supports IPv4 and IPv6 loopback, TLS 1.2 and TLS 1.3, classical and hybrid configurations, mTLS, PSK resumption, HRR, and chain-depth variation. All execution parameters are specified by the artifact configuration.

\subsection{Baseline execution}

\Bzero is executed as a Linux/container baseline when the reference implementation supports the scenario. Its outputs are normalized into the same comparative frame as \Bone, \Btwo, and \Bthree. We treat \Bzero as an inherited detector with narrower semantics than the multi-surface measurement system. The comparison therefore reports detection behavior and plane closure on a shared scenario slice.

For the public scanner comparison, we run SSLyze and testssl.sh on a 250-target stratified slice. We compare their outputs with single classical probe, single hybrid probe, dual-probe aggregation, and dual-probe-plus-chain modes. All tools are evaluated under the configured target set, probe budget, and campaign policy.

\subsection{Stratified public campaign profiles and guardrails}

The public campaign measures 1000 public targets across 10 operational families, with 100 targets per family: API endpoints, CDN edges, cloud vendors, commerce, developer documentation, government, knowledge communities, media/news, package ecosystems, and universities. Each target receives two TLS 1.3 probe profiles, producing 2000 probes per round.

The classical profile offers X25519 as its supported group. The hybrid-capable profile offers X25519 and \hybrid. Both profiles use the same signature-scheme set, \texttt{rsa\_pss\_rsae\_sha256} and \texttt{ecdsa\_secp256r1\_sha256}. Client authentication is disabled in both profiles. This profile pair isolates the effect of supported-group choice while holding authentication offers fixed.

The campaign policy requires explicit SNI for DNS hostnames, uses HTTPS/443 targets, disables client-auth profiles for public targets, permits at most one retry per probe, applies a fixed 30-second backoff policy, and caps concurrency at 20 workers. The inventory records target family, owner scope, measurement tier, profile identifiers, selection source, and selection basis. Public campaign results are scoped to the declared target set, strata, probe profiles, and vantage conditions.

\begin{table}[t]
	\centering
	\caption{Public campaign probe profiles. Both profiles use TLS 1.3 and the same signature-scheme set.}
	\label{tab:public-profiles}
	\small
	\begin{tabular}{@{}ll@{}}
		\toprule
		\textbf{Profile} & \textbf{Supported groups} \\
		\midrule
		Classical default & X25519 \\
		Hybrid-capable default & X25519, \hybrid \\
		\bottomrule
	\end{tabular}
\end{table}

\subsection{Repeated campaign rounds}

The campaign plan supports up to five rounds over the same inventory, with 2000 probes per round. R0 is the main public campaign round used for the primary public measurement results. R1 repeats the same 1000-target, 2000-probe design 48 hours later. We use R1 to assess short-term stability in the R0 measurement pattern.

\subsection{Statistical treatment}

For family-level proportions, including hybrid-confirmed rate, we report Wilson confidence intervals. For paired comparisons of hybrid discovery between single-probe and dual-probe modes, we use paired tests such as McNemar's test. These tests quantify observed differences within the measured slice.

\section{Controlled Benchmark Results}
\label{sec:benchmark-results}

\subsection{Canonical plane closure}

\Cref{tab:canonical-plane-closure} reports canonical suite results.

\begin{table*}[t]
	\centering
	\caption{Canonical-suite results over 14 executed scenarios. Plane closure is reported as a fraction of scenarios.}
	\label{tab:canonical-plane-closure}
	\small
	\setlength{\tabcolsep}{3.5pt}
	\begin{tabular}{@{}l *{11}{S[table-format=1.2]}@{}}
		\toprule
		Mode &
		\multicolumn{1}{c}{\makecell{Session\\core}} &
		\multicolumn{1}{c}{\makecell{Hidden\\detail}} &
		\multicolumn{1}{c}{\makecell{Capability}} &
		\multicolumn{1}{c}{\makecell{Key\\est.}} &
		\multicolumn{1}{c}{\makecell{Auth}} &
		\multicolumn{1}{c}{\makecell{Lifecycle}} &
		\multicolumn{1}{c}{\makecell{Observ.}} &
		\multicolumn{1}{c}{\makecell{Obj.\\comp.}} &
		\multicolumn{1}{c}{\makecell{Obj.\\clear}} &
		\multicolumn{1}{c}{\makecell{Primary\\score}} &
		\multicolumn{1}{c}{\makecell{Ambig.}} \\
		\midrule
		B1 passive-only    & 1.00 & 0.29 & 0.00 & 1.00 & 0.29 & 0.29 & 1.00 & 0.00 & 0.00 & 0.35 & 1.00 \\
		B2 passive-active  & 1.00 & 1.00 & 1.00 & 1.00 & 0.29 & 0.29 & 1.00 & 0.29 & 0.29 & 0.76 & 0.71 \\
		B3 multi-surface   & 1.00 & 1.00 & 1.00 & 1.00 & 1.00 & 1.00 & 1.00 & 1.00 & 1.00 & 1.00 & 0.00 \\
		\bottomrule
	\end{tabular}
\end{table*}

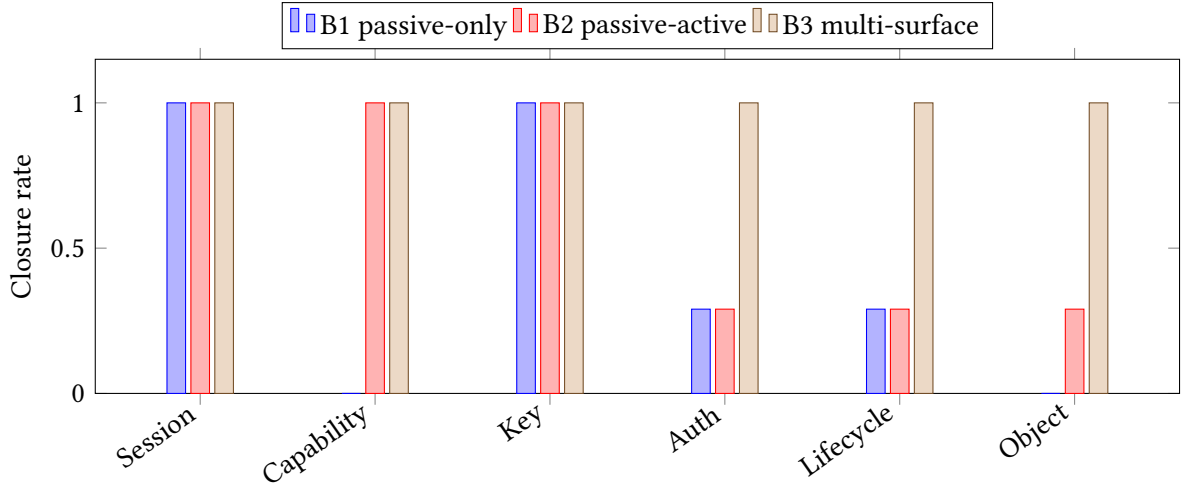
\begin{figure}[t]
	\centering
	\begin{tikzpicture}
\begin{axis}[
    ybar,
    width=\textwidth,
    height=6.0cm,
    ymin=0, ymax=1.15,
    ylabel={Closure rate},
    symbolic x coords={Session,Capability,Key,Auth,Lifecycle,Object},
    xtick=data,
    x tick label style={rotate=35,anchor=east},
    legend style={at={(0.5,1.03)},anchor=south,legend columns=3},
    bar width=7pt,
    enlarge x limits=0.12,
]
\addplot coordinates {(Session,1.00) (Capability,0.00) (Key,1.00) (Auth,0.29) (Lifecycle,0.29) (Object,0.00)};
\addplot coordinates {(Session,1.00) (Capability,1.00) (Key,1.00) (Auth,0.29) (Lifecycle,0.29) (Object,0.29)};
\addplot coordinates {(Session,1.00) (Capability,1.00) (Key,1.00) (Auth,1.00) (Lifecycle,1.00) (Object,1.00)};
\legend{B1 passive-only,B2 passive-active,B3 multi-surface}
\end{axis}
\end{tikzpicture}
	\caption{Canonical plane closure by measurement mode. Passive-only evidence closes session and key establishment. Active evidence closes capability. Certificate-chain evidence closes authentication and lifecycle.}
	\label{fig:canonical-plane-closure}
\end{figure}

The passive-only mode closes the session and key-establishment planes on the canonical suite. It also closes authentication and lifecycle when the protocol exposes certificate material passively, which in this benchmark corresponds to the TLS 1.2 cases. Its primary score and primary coverage reflect this evidence boundary, and ambiguity remains high in planes that require active or chain evidence.

The passive-active mode closes the capability plane. It records what controlled profiles confirm the endpoint can negotiate and keeps that capability evidence separate from the parameters negotiated in a captured session. Authentication and lifecycle remain open without chain evidence, except in TLS 1.2 cases where certificate material is passively visible.

The multi-surface mode closes all canonical planes. When the benchmark supplies the relevant evidence surfaces and linkage, the measurement object reaches full canonical closure while preserving separate session, capability, authentication, and lifecycle profiles.

\subsection{Stress-contract behavior}

The stress suite is evaluated by behavioral contract. The contracts test uncertainty preservation, contradiction detection, and correct use of \na.

\begin{table}[t]
	\centering
	\caption{Stress-contract evaluation over 15 scenarios. Stress scenarios reward correct uncertainty and contradiction preservation rather than field completion.}
	\label{tab:stress-contract-summary}
	\small
	\begin{tabular}{@{}lrrr@{}}
		\toprule
		\textbf{Metric} & \textbf{\Bone} & \textbf{\Btwo} & \textbf{\Bthree} \\
		\midrule
		Contract match & 0.99 & 1.00 & 1.00 \\
		Contract coverage & 1.00 & 1.00 & 1.00 \\
		Ambiguity rate & 1.00 & 0.93 & 0.33 \\
		Contradiction rate & 0.00 & 0.20 & 0.20 \\
		Capability broader than negotiation & 0.00 & 0.07 & 0.07 \\
		Cross-surface/temporal contradiction & 0.00 & 0.20 & 0.20 \\
		\midrule
		Session core & 0.87 & 0.87 & 0.87 \\
		Session hidden detail & 0.13 & 0.93 & 0.93 \\
		Capability & 0.00 & 1.00 & 1.00 \\
		Key establishment & 0.93 & 0.93 & 0.93 \\
		Authentication & 0.13 & 0.13 & 1.00 \\
		Lifecycle & 0.13 & 0.13 & 1.00 \\
		Observability & 0.87 & 0.67 & 0.67 \\
		Object completeness & 0.00 & 0.13 & 0.87 \\
		Object complete clear & 0.00 & 0.07 & 0.67 \\
		\bottomrule
	\end{tabular}
\end{table}

The stress suite shows three main behaviors. First, incomplete passive evidence reduces session closure as expected: session core closes in 0.87 of stress scenarios. Second, active probing closes the capability plane and exposes contradiction-bearing evidence in scenarios designed for cross-surface disagreement. Third, multi-surface evidence raises authentication and lifecycle closure to 1.00 while preserving contradiction in 0.20 of stress scenarios. In these cases, correctness is measured by evidence-faithful state assignment, including explicit uncertainty, contradiction, and \na where the scenario semantics require them.

\subsection{Inherited baseline comparison}

\Cref{tab:b0-comparison} compares \Bzero, \Bone, \Btwo, and \Bthree on the 29-scenario shared local benchmark slice.

\begin{table}[t]
	\centering
	\caption{Shared local benchmark-slice comparison over 29 executed scenarios.}
	\label{tab:b0-comparison}
	\small
	\begin{tabular}{@{}lrr@{}}
		\toprule
		\textbf{Reported property} & \textbf{Overall} & \textbf{TLS 1.3} \\
		\midrule
		\Bzero detected run & 2/29 & 0/23 \\
		\Bone session-core closure & 27/29 & 21/23 \\
		\Bone session-hidden-detail closure & 6/29 & -- \\
		\Btwo capability closure & 29/29 & 23/23 \\
		\Bthree object completeness & 27/29 & 21/23 \\
		\Bthree object complete and clear & 24/29 & 19/23 \\
		\bottomrule
	\end{tabular}
\end{table}

The inherited baseline detects 2 of 29 runs and 0 of 23 TLS 1.3 runs on this local slice under our benchmark execution and normalization regime. Its missed detections concentrate in TLS 1.3-heavy and stress-contract scenarios, where encrypted authentication, missing linkage, hidden detail, and unresolved evidence are part of the expected measurement object. The comparison shows the difference between packet-level detection and multi-surface measurement: the former reports selected visible indicators, whereas the latter assigns evidence to planes, records linkage, and preserves unresolved or contradictory states.

The distinction between object completeness and clear completeness is also relevant. A result can be structurally complete and still contain a contradiction. Clear completeness requires plane closure and contradiction-free linkage.

\section{Stratified Public Measurement Study}
\label{sec:public-study}

\subsection{R0 campaign overview}

The first public campaign round, R0, covers 1000 public targets with 2000 fresh probes. \Cref{tab:r0-summary} summarizes the outcome.

\begin{table}[t]
\centering
\caption{R0 stratified public campaign summary.}
\label{tab:r0-summary}
\begin{tabular}{@{}l r@{}}
\toprule
Metric & Count \\
\midrule
Public targets & 1000 \\
Fresh probes & 2000 \\
Complete handshakes & 1971 \\
Certificate-chain artifacts & 1368 \\
Hybrid-confirmed targets & 310 \\
Classical-only targets under tested profiles & 690 \\
Capability broader than single-session view & 310 \\
Contradiction-bearing targets & 2 \\
Dual-probe-plus-chain clear-complete targets & 682 \\
\bottomrule
\end{tabular}
\end{table}

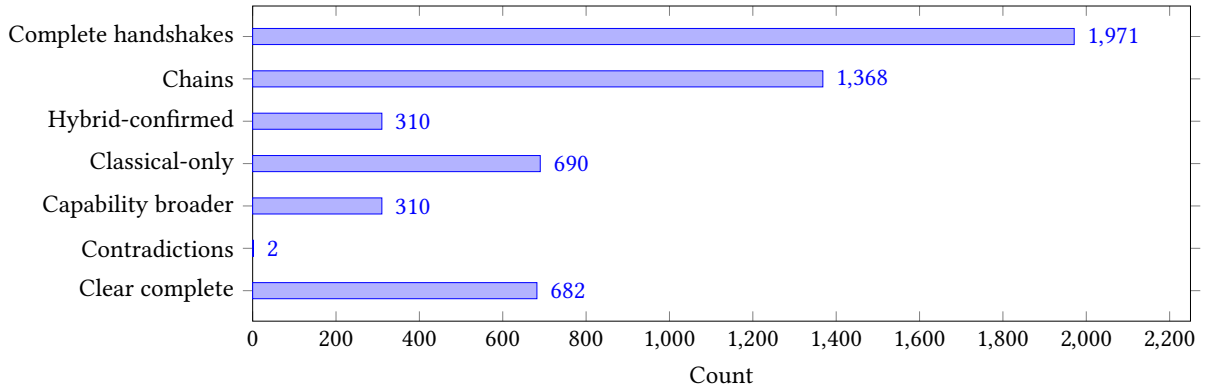
\begin{figure}[t]
	\centering
		\resizebox{\textwidth}{!}{%
		\begin{tikzpicture}
			\begin{axis}[
				xbar,
				width=\textwidth,
				height=6.4cm,
				xmin=0,
				xmax=2250,
				xlabel={Count},
				symbolic y coords={
					Clear complete,
					Contradictions,
					Capability broader,
					Classical-only,
					Hybrid-confirmed,
					Chains,
					Complete handshakes
				},
				ytick=data,
				nodes near coords,
				nodes near coords style={
					anchor=west,
					xshift=2pt
				},
				enlarge y limits=0.12,
				bar width=7pt,
				]
				\addplot coordinates {
					(682,Clear complete)
					(2,Contradictions)
					(310,Capability broader)
					(690,Classical-only)
					(310,Hybrid-confirmed)
					(1368,Chains)
					(1971,Complete handshakes)
				};
			\end{axis}
		\end{tikzpicture}%
	}
	\caption{R0 public campaign summary. Counts are target-level except for probes, complete handshakes, and chain artifacts.}
	\label{fig:ccs-r0-summary}
\end{figure}

The campaign confirms hybrid capability for 310 targets. These same 310 targets are also capability-broader-than-single-session cases: hybrid support appears under deliberate multiprobe measurement, while a single classical session view exposes only classical negotiation. The campaign records 2 contradiction-bearing targets, showing cross-surface disagreement in the public measurement setting.

The dual-probe-plus-chain mode reaches clear completeness for 682 targets. Under the declared profiles and chain collection policy, these targets have cleanly closed measurement objects across the required planes.

\subsection{Repeated round and temporal stability}

A second full campaign round, R1, repeats the 1000-target, 2000-probe measurement. R1 reproduces the main R0 capability pattern: hybrid-confirmed targets remain 310, classical-only-under-tested-profiles targets remain 690, and capability-broader-than-session targets remain 310. Clear-complete targets decrease from 682 to 680, while contradiction-bearing targets increase from 2 to 4.

\begin{table}[t]
	\centering
	\caption{Repeated public campaign rounds. R1 confirms the R0 hybrid-capability result with minor drift.}
	\label{tab:r0-r1}
	\small
	\begin{tabular}{@{}lrrr@{}}
		\toprule
		\textbf{Metric} & \textbf{R0} & \textbf{R1} & \textbf{$\Delta$} \\
		\midrule
		Targets & 1000 & 1000 & 0 \\
		Probes & 2000 & 2000 & 0 \\
		Complete handshakes & 1971 & 1968 & -3 \\
		Chain artifacts & 1368 & 1368 & 0 \\
		Hybrid-confirmed targets & 310 & 310 & 0 \\
		Classical-only targets & 690 & 690 & 0 \\
		Capability broader than session & 310 & 310 & 0 \\
		Clear-complete targets & 682 & 680 & -2 \\
		Contradiction-bearing targets & 2 & 4 & +2 \\
		\bottomrule
	\end{tabular}
\end{table}

For round-level drift, the artifact compares targets with comparable multi-surface records in both rounds. On the 684 targets in this comparison, capability drift is 0.29\%, certificate drift is 3.95\%, lifecycle-bucket drift is 0.00\%, signature-algorithm drift is 0.15\%, signature-scheme drift is 0.00\%, and clear-complete stability is 99.42\%. These results support a short-term temporal stability claim for the R0 measurement pattern.

\subsection{Family heterogeneity}

The campaign is stratified by operational family. \Cref{tab:family-wilson} reports selected family results with Wilson intervals. The full artifact contains the complete family table.

\begin{table}[t]
\centering
\caption{Selected family-level hybrid-confirmed rates in R0 with Wilson 95\% intervals.}
\label{tab:family-wilson}
\begin{tabular}{@{}l c c@{}}
\toprule
Family & Hybrid-confirmed rate & Wilson 95\% interval \\
\midrule
Knowledge community & 0.620 & [0.522, 0.709] \\
Government & 0.110 & [0.063, 0.186] \\
Cloud vendor & 0.110 & [0.063, 0.186] \\
\bottomrule
\end{tabular}
\end{table}

Hybrid-confirmed rates vary substantially by operational family. The knowledge-community family has a hybrid-confirmed rate of 0.620 with a 95\% Wilson interval of [0.522, 0.709]. Government and cloud-vendor families each have rates of 0.110 with intervals [0.063, 0.186]. These differences show that target family materially affects the hybrid capability observed by the campaign.

\subsection{Statistical comparison of probing modes}

The baseline-zoo slice compares single classical probing with dual probing on paired targets. The single classical probe identifies 0 hybrid-capable targets, while dual probing confirms hybrid capability for 70 targets. The paired comparison has 70 discordant pairs in favor of dual probing and 0 in the opposite direction, yielding a McNemar p-value of $2.44\times 10^{-13}$. A bootstrap estimate of hybrid-discovery uplift against the single classical probe is 0.28, with 95\% interval [0.224, 0.336]. In the measured slice, hybrid discovery is strongly associated with probing mode.

\section{Baseline Zoo Comparison}
\label{sec:baseline-zoo}

\subsection{Scanner and probe comparison}

\Cref{tab:baseline-zoo} compares SSLyze, testssl.sh, single-probe modes, dual-probe aggregation, and dual-probe-plus-chain measurement on a 250-target stratified slice.

\begin{table*}[t]
	\centering
	\caption{Baseline-zoo comparison on a 250-target stratified slice.}
	\label{tab:baseline-zoo}
	\small
	\setlength{\tabcolsep}{4pt}
	\begin{tabularx}{\textwidth}{@{}l c c c Y@{}}
		\toprule
		Mode/tool &
		\makecell{Session\\signal} &
		\makecell{Certificate\\signal} &
		\makecell{Hybrid\\support} &
		Notes \\
		\midrule
		SSLyze & 192/250 & 192/250 & 0/250 & Strong classical session/certificate signal under budget. \\
		testssl.sh & 42/250 & 40/250 & 0/250 & 183 timeouts and 25 errors under budget. \\
		Single classical probe & 177/250 & -- & 0/250 & Classical profile only. \\
		Single hybrid probe & 177/250 & -- & 70/250 & Hybrid-capable profile. \\
		Dual-probe aggregated & 174/250 & -- & 70/250 & Capability aggregation across profiles. \\
		Dual-probe + chain & 174/250 & 174/250 & 70/250 & 173/250 clear-complete objects. \\
		\bottomrule
	\end{tabularx}
\end{table*}

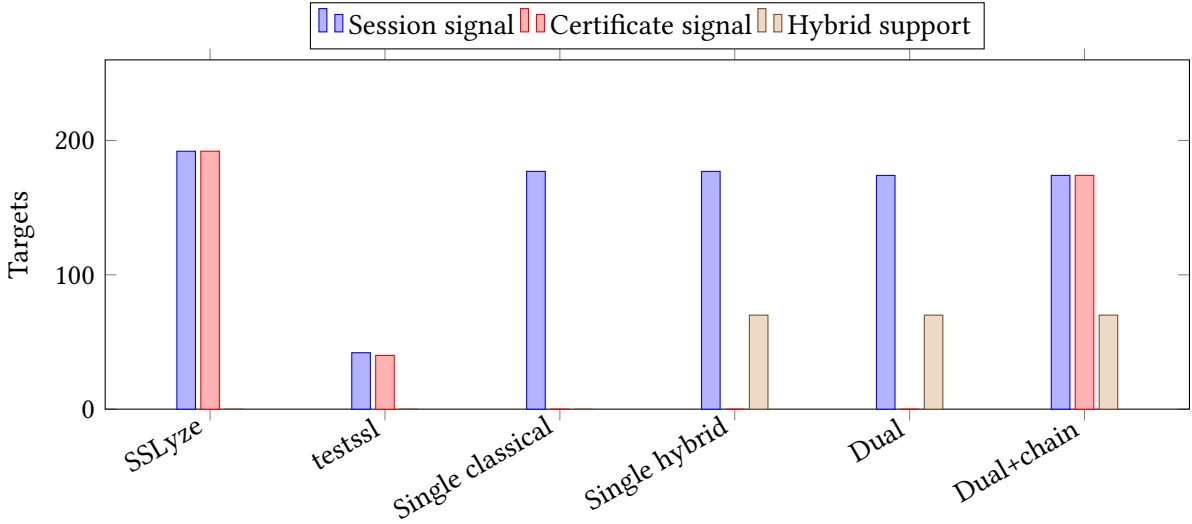
\begin{figure}[t]
	\centering
	\begin{tikzpicture}
\begin{axis}[
    ybar,
    width=\textwidth,
    height=6.2cm,
    ymin=0, ymax=260,
    ylabel={Targets},
    symbolic x coords={SSLyze,testssl,Single classical,Single hybrid,Dual,Dual+chain},
    xtick=data,
    x tick label style={rotate=30,anchor=east},
    legend style={at={(0.5,1.03)},anchor=south,legend columns=3},
    bar width=7pt,
    enlarge x limits=0.12,
]
\addplot coordinates {(SSLyze,192) (testssl,42) (Single classical,177) (Single hybrid,177) (Dual,174) (Dual+chain,174)};
\addplot coordinates {(SSLyze,192) (testssl,40) (Single classical,0) (Single hybrid,0) (Dual,0) (Dual+chain,174)};
\addplot coordinates {(SSLyze,0) (testssl,0) (Single classical,0) (Single hybrid,70) (Dual,70) (Dual+chain,70)};
\legend{Session signal,Certificate signal,Hybrid support}
\end{axis}
\end{tikzpicture}
	\caption{Baseline-zoo comparison on the 250-target stratified slice. Classical scanners provide session and certificate signals, while hybrid support appears under hybrid-capable or dual-probe modes in this slice.}
	\label{fig:baseline-zoo}
\end{figure}

The comparison shows three patterns. First, testssl.sh has low effective coverage under the configured budget, with 183 timeouts and 25 error or unresolved outcomes. This result is scoped to the configured slice and execution budget. Second, SSLyze provides broader session and certificate signal coverage. Third, the scanner rows report 0 hybrid-confirmed targets in this slice, while the hybrid-capable and dual-probe modes confirm 70 hybrid-capable targets. Adding chain evidence closes 173 measurement objects cleanly.

\subsection{Family-level scanner behavior}

\Cref{tab:baseline-family-examples} gives selected family examples from the same 250-target slice.

\begin{table}[t]
\centering
\caption{Selected family-level baseline-zoo counts on the 250-target slice.}
\label{tab:baseline-family-examples}
\begin{tabular}{@{}l r r r r@{}}
\toprule
Family & testssl signal & SSLyze signal & Dual hybrid & Dual+chain clear \\
\midrule
API endpoint & 9 & 22 & 13 & 20 \\
Government & 0 & 25 & 1 & 12 \\
Knowledge community & 0 & 24 & 16 & 21 \\
\bottomrule
\end{tabular}
\end{table}

Family-level results show heterogeneous behavior across operational categories. In the knowledge-community family, SSLyze produces 24 signals and dual-probe measurement confirms 16 hybrid targets, while testssl.sh produces 0 signals under the configured budget. In the government family, SSLyze produces 25 signals and dual-probe measurement confirms 1 hybrid target. The visible measurement outcome therefore depends on both methodology and target family.

\section{Discussion}
\label{sec:discussion}

\subsection{Support and negotiation are separate claims}

The public campaign and the support-versus-negotiation stress scenarios show that endpoint capability requires active profile variation. A classical client profile may negotiate a classical group with an endpoint that negotiates X25519MLKEM768 under a hybrid-capable profile. Readiness for hybrid-client interoperability is therefore supported by active capability evidence. The cryptographic properties of a captured flow are supported by session evidence. These two claims belong to different measurement planes.

This separation is central for crypto-agility dashboards. A dashboard based only on the last observed session can miss latent hybrid capability. A dashboard based only on supported groups can overstate protection for current traffic. A faithful dashboard should report a vector: observed session behavior, confirmed capability lower bound, authentication profile, lifecycle profile, and linkage confidence.

\subsection{Certificate evidence requires linkage}

TLS 1.3 encrypts certificates in the handshake, so active chain collection is operationally useful and should be recorded as active endpoint evidence. A chain retrieved after a session closes endpoint-authentication evidence for the probe time. Its relationship to a previously captured session depends on timing, routing, server configuration, and certificate deployment. The artifact represents this relationship through chain source types and linkage strength. Temporal-drift scenarios show how passive session evidence from time $t$ and chain evidence from time $t+\Delta$ can disagree across surfaces.

\subsection{Uncertainty as measured output}

Traditional accuracy metrics reward populated outputs. Observability benchmarks also need to reward faithful unresolved states. In a truncated capture, an \unknown selected group is the correct result. In PSK resumption, fresh authentication may be unavailable, inherited, or non-applicable. In TLS 1.3 mTLS, hidden client-auth detail remains unresolved in passive capture unless linked same-run active evidence closes it. Stress contracts encode these cases so that tools are scored for preserving \unknown, \na, ambiguity, and contradiction when the evidence requires them.

\subsection{Scanner baselines as partial evidence}

SSLyze and testssl.sh remain useful operational tools. The baseline-zoo comparison shows their role as partial evidence sources for post-quantum TLS observability. Under the configured slice, their hybrid-confirmed count is 0, while deliberate dual-probe measurement confirms 70 hybrid-capable targets. Post-quantum readiness assessment therefore benefits from probe profiles designed to elicit hybrid behavior and rules that keep capability separate from negotiation.

\subsection{Answers to the research questions}

\begin{table*}[t]
	\centering
	\caption{Summary answers to the research questions.}
	\label{tab:rq-answers}
	\small
	\begin{tabularx}{\textwidth}{@{}p{0.08\textwidth}X@{}}
		\toprule
		\textbf{RQ} & \textbf{Answer} \\
		\midrule
		RQ1 & Passive evidence closes session and key-establishment planes in complete canonical traces. Endpoint capability requires active evidence, and TLS 1.3 authentication usually requires chain evidence. In the canonical suite, passive-only closes session core and key establishment at 1.00, capability at 0.00, and authentication/lifecycle at 0.29. \\
		RQ2 & Active probing closes capability lower bounds and reveals capability-broader-than-session cases. In R0, 310 targets are hybrid-confirmed and capability-broader-than-single-session. \\
		RQ3 & Certificate-chain evidence closes authentication and lifecycle when source, timestamp, and linkage are explicit. TLS 1.3 active chain collection supplies endpoint evidence with probe-time scope. \\
		RQ4 & Incomplete and contradictory cases are represented directly. Stress contracts show that \unknown, \na, ambiguity, and contradiction are correct outputs for selected scenario semantics. \\
		RQ5 & Under the 250-target slice and configured budget, SSLyze and testssl.sh provide useful classical signals and report 0 hybrid-confirmed targets, while dual-probe measurement confirms 70 hybrid-capable targets. \\
		\bottomrule
	\end{tabularx}
\end{table*}

\subsection{Operator guidance}

The findings suggest the following operational guidance:

\begin{enumerate}[leftmargin=1.4em]
	\item Use passive evidence to audit observed sessions, including negotiated TLS version, selected group when visible, HRR, and capture completeness.
	\item Use active probes to estimate lower bounds on endpoint capability and keep those results separate from captured-session claims.
	\item Use chain collection to close authentication and lifecycle, with source, timestamp, and linkage recorded alongside the result.
	\item Use dual or multi-profile probing when assessing hybrid support. A single classical profile can miss hybrid capability.
	\item Treat \unknown, \na, ambiguity, and contradiction as first-class outputs.
	\item Report readiness as a vector with separate key-establishment, authentication, lifecycle, and evidence-confidence components.
\end{enumerate}

\section{Related Work}
\label{sec:related}

\paragraph{TLS measurement and scanning.}
TLS measurement has a long history in Internet security research, including studies of protocol versions, cipher suites, certificate chains, downgrade risks, and misconfiguration. Operational scanners such as SSLyze and testssl.sh provide practical visibility into many TLS properties~\cite{SSLyze,testssl}. Our work focuses on post-quantum TLS observability and separates support, negotiation, authentication, lifecycle, and linkage.

\paragraph{Post-quantum TLS.}
Hybrid TLS key exchange and ML-DSA authentication are active areas of standardization and deployment~\cite{IETFECDHEMLKEM,IETFMLDSA,IETFUTAPQC}. Industry deployments and browser experiments have demonstrated both the feasibility and operational complexity of hybrid key exchange~\cite{ChromiumKyber,CloudflarePQC}. Our work measures how these constructions appear across passive, active, certificate-chain, and registry surfaces.

\paragraph{Crypto-agility and inventory.}
Crypto-agility guidance emphasizes the identification of cryptographic assets, algorithms, dependencies, and deployment contexts before migration~\cite{NISTCryptoAgility2025}. This paper operationalizes that principle for TLS by defining evidence surfaces, measurement planes, and provenance-aware inference rules.

\paragraph{Quantum-vulnerable TLS packet inspection.}
Cho et al. propose automated packet inspection for quantum-vulnerable TLS, including hybrid TLS detection and active certificate retrieval for TLS 1.3~\cite{Cho2026CryptoAgilityTLS}. Our work extends this line with a benchmark, an evidence model, and an evaluation methodology that assigns each observation to the surface and plane it supports.

\section{Limitations and Threats to Validity}
\label{sec:limitations}

\paragraph{Public campaign representativeness.}
The 1000-target public campaign is stratified and operationally diverse. Its family-level rates characterize the measured inventory, target-selection procedure, probe profiles, and vantage conditions. They support stratified measurement claims for this campaign.

\paragraph{Vantage point and client profiles.}
Active probing results depend on vantage point, client profile, supported groups, timeout budget, and server behavior. A target classified as classical-only-under-tested-profiles may expose hybrid behavior under another profile or path. We therefore report capability as a confirmed lower bound under declared profiles.

\paragraph{Repeated-round interpretation.}
R1 provides a short-term repeat of the public campaign over the same inventory and profiles. Its role is to assess immediate stability of the R0 measurement pattern. The low capability drift and high clear-complete stability strengthen the R0 result within this short-term window.

\paragraph{Certificate-chain linkage.}
TLS 1.3 chain evidence collected through active probing describes the endpoint at probe time. Its linkage to a captured flow depends on timing, routing, server configuration, and certificate deployment. The artifact records linkage strength and contradiction so that temporal drift remains visible in the measurement object.

\paragraph{Baseline comparability.}
\Bzero, SSLyze, and testssl.sh have different goals, inputs, and output semantics. We compare them on observable signals, hybrid-confirmed counts, and closure properties under a shared target slice and execution budget. This comparison reflects the operational setting in which such tools may be used for TLS readiness auditing.

\paragraph{Coverage of PKI complexity.}
The benchmark covers selected PKI features, including chain-depth variation and public campaign chain collection. Further work can extend the corpus with mixed roots, cross-signatures, delegated credentials, and future post-quantum certificate formats. The current benchmark provides a foundation for evaluating certificate-chain evidence within the broader multi-surface model.

\paragraph{Evolving standards.}
Post-quantum TLS identifiers and drafts continue to evolve. The versioned registry records identifier status, aliases, and provenance so that new standards can be incorporated while preserving historical deployment distinctions. The core methodology depends on the separation between evidence surfaces and measurement planes.

\paragraph{Ethics and measurement impact.}
The public campaign uses active probes against public endpoints. The artifact includes inventory guardrails, separates public and cooperative targets, and excludes client-auth probes for blind public targets. Active measurement should use conservative rates, clear identification where appropriate, and respect for institutional policies.

\section{Reproducibility and Artifact}
\label{sec:artifact}

The artifact is organized around versioned registries, JSON Schemas, scenario manifests, observation records, inference rules, stress contracts, evaluation scripts, baseline adapters, campaign manifests, verifier scripts, and paper-ready exports. The central benchmark release contains 29 controlled scenarios. The public campaign layer records target inventories, probe manifests, reused and fresh probe accounting, chain artifacts, baseline-zoo outputs, repeated-round summaries, drift analysis, and statistical tests.

The repository separates measurement code from measurement semantics. Parsers extract raw and canonical evidence. Registries normalize identifiers. Inference rules construct measurement objects. Policy profiles are applied downstream. Stress contracts define expected behavior for incomplete, contradictory, or non-observable cases. This organization makes the core measurement invariant auditable: active evidence can enrich a measurement object while retaining its active source label.

The artifact includes verifier scripts for the frozen benchmark release, the inherited \Bzero comparison, the CCS public-campaign extension, and the baseline-zoo slice. The freeze manifest records checksums for benchmark manifests, canonical and stress summaries, baseline comparison exports, longitudinal overview exports, and paper-ready tables.

\section{Conclusion}
\label{sec:conclusion}

Post-quantum TLS migration is measured through structured evidence about sessions, endpoints, certificate chains, and identifier semantics. A captured session exposes negotiated behavior. Active probes elicit endpoint capability under declared profiles. Certificate chains express authentication and lifecycle properties. Registries map raw protocol identifiers into cryptographic meaning. These surfaces support different claims and require explicit provenance.

This paper presented a multi-surface measurement framework for post-quantum TLS observability, together with a schema-enforced reproducible artifact and a 29-scenario benchmark covering canonical and stress cases across TLS 1.2, TLS 1.3, classical and hybrid key establishment, mTLS, PSK resumption, HelloRetryRequest, truncation, fragmentation/coalescing, temporal drift, IPv6, and certificate-chain variation. The controlled evaluation shows how each evidence surface closes different planes: passive evidence closes session-level planes, active probing closes capability lower bounds, and certificate-chain evidence closes authentication and lifecycle when source and linkage are explicit. The inherited packet-inspection baseline detected 2 of 29 local runs and 0 of 23 TLS 1.3 runs, illustrating the gap between visible packet-level indicators and full measurement-object closure.

The public campaign shows that measurement mode materially changes observed post-quantum readiness. Across 1000 targets and 2000 fresh probes in R0, the framework completed 1971 handshakes, collected 1368 chain artifacts, and confirmed hybrid capability for 310 targets. Those same 310 targets exhibited capability broader than a single classical session view, showing that hybrid readiness can require profile variation to become visible. A repeated R1 round preserved the same hybrid-confirmed count and showed high clear-complete stability, while surfacing small capability and certificate drift. On the scanner comparison slice, general-purpose TLS tools provided useful classical session and certificate signals under the configured budget, while dual-probe measurement confirmed hybrid support.

The central result is an operational measurement discipline for post-quantum TLS auditing. Readiness should be reported as a structured measurement object with separate session behavior, endpoint capability, authentication, lifecycle, linkage, and policy components. The same object should preserve \unknown, \na, ambiguity, and contradiction as first-class outcomes. Crypto-agility auditing depends on this discipline: incomplete evidence should remain visible until the corresponding measurement plane is closed by an appropriate surface.

\newpage

\printbibliography

\newpage

\section{Appendices}

\appendix
\section{Formal Measurement Object}
\label{app:measurement-object}

The artifact represents each inferred result as a structured object, a simplified view is shown below.

\begin{lstlisting}[basicstyle=\ttfamily\small,breaklines=true]
{
  "scenario_id": "...",
  "mode": "multi_surface",
  "session_profile": {
    "negotiated_tls_version": "TLS1.3",
    "selected_group": "X25519MLKEM768",
    "hrr_seen": true,
    "completeness_status": "complete",
    "mtls_seen": "unknown_or_linked_true"
  },
  "key_establishment_profile": {
    "profile": "hybrid",
    "components": ["X25519", "ML-KEM-768"],
    "applicability_state": "applicable"
  },
  "capability_profile": {
    "probe_profiles": ["classical", "hybrid"],
    "supported_groups_lower_bound": ["X25519", "X25519MLKEM768"],
    "capability_broader_than_session": true
  },
  "authentication_profile": {
    "leaf_spki_algorithm": "ECDSA",
    "leaf_signature_algorithm": "ecdsa-with-SHA256",
    "chain_depth": 2,
    "chain_source_type": "active_probe"
  },
  "lifecycle_profile": {
    "validity_days": 90,
    "short_lived_bucket": true
  },
  "observability_profile": {
    "surface_origins": ["SigmaP", "SigmaA", "SigmaC"],
    "plane_linkage": {"session": "passive", "authentication": "active_chain"},
    "contradiction_flag": false,
    "ambiguity_reasons": []
  }
}
\end{lstlisting}

This representation allows the evaluation to distinguish a complete and clear object from an object that is structurally complete but contradiction-bearing. It also allows policy projections to change without rewriting the underlying measurement.

\section{Benchmark Scenario Families}
\label{app:benchmark-families}

The canonical suite includes clean executions for the following families: TLS 1.2 static RSA, TLS 1.2 ECDHE\_RSA, TLS 1.2 ECDHE\_ECDSA, TLS 1.2 mTLS, TLS 1.3 X25519, TLS 1.3 X25519MLKEM768, RSA and ECDSA leaf authentication, TLS 1.3 mTLS sentinels, certificate lifecycle variants, IPv6 loopback, and leaf-plus-intermediate chain depth.

The stress suite includes support-vs-negotiation divergence, truncated passive captures, fragmentation/coalescing, temporal drift, HelloRetryRequest with degraded passive evidence, PSK/resumption continuity, and mTLS/resumption crossover. Each stress scenario is associated with a behavioral contract. The contract states which planes should close, which should remain unknown, which contradictions should be detected, and which properties are not applicable.

\section{Stress Contract Examples}
\label{app:stress-contracts}

Stress contracts are needed because some scenarios make exact-match labels inappropriate. The following examples illustrate the contract style.

\paragraph{Pre-ServerHello truncation.} If the passive capture ends before the relevant server negotiation, the selected group must remain unresolved. A tool that guesses the configured expected group from external knowledge fails the contract because it over-resolves beyond the passive surface.

\paragraph{Support versus negotiation.} If a classical probe negotiates X25519 and a hybrid-capable probe negotiates X25519MLKEM768 against the same endpoint, the endpoint capability lower bound includes both groups. The observed classical session remains classical. The contract requires both facts to be preserved.

\paragraph{Temporal drift.} If passive evidence and later active or chain evidence disagree, the contract expects a contradiction marker. The contradiction is not necessarily an error in the parser, it is instead a property of the measurement interval.

\paragraph{TLS 1.3 mTLS hidden detail.} Passive absence of a visible client certificate request is not evidence that mTLS was not used. The contract expects \texttt{unknown} or linked same-run active evidence, not passive \texttt{false}.

\end{document}